\documentclass{article}
\usepackage[top=2cm,right=2cm,bottom=3cm,left=2cm]{geometry}

\usepackage{graphicx}

\usepackage{amssymb,amsthm,amsmath,bm,cite}
\usepackage[nonindentfirst]{titlesec}
\usepackage[ruled]{algorithm2e}
\usepackage{lipsum}
\theoremstyle{definition}

\usepackage[symbol]{footmisc}

\makeatletter\renewcommand\maketitle{
{\begin{center}
{\Large\@title}
\vspace{1em}\\
{\@author}
\vspace{1em}
\end{center}}}\makeatother

\titleformat{\section}
{\small\bfseries}
{\thesection.}{0.2em}{}
\titlespacing{\section}
{0pt}{1.5em}{0.5em}

\titleformat{\subsection}
{\small\itshape}
{\thesubsection.}{0.2em}{}
\titlespacing{\section}
{0pt}{1.5em}{0.5em}

\titleformat{\subsubsection}
{\small\itshape}
{\thesubsubsection.}{0.2em}{}
\titlespacing{\section}
{0pt}{1.5em}{0.5em}

\providecommand{\keywords}[1]{
  \small	
  \hspace{10pt}\textbf{Keywords---}\textit{#1}
}

\usepackage{fancyhdr}
\begin{document}
\thispagestyle{fancy}
\title{From ductile damage to unilateral contact via point-wise implicit discontinuity at the infinitesimal element level}
\author{Alireza Daneshyar\footnote[1]{Corresponding author.\newline\hspace*{15pt} E-mail: alireza.daneshyar@tum.de}, Leon Herrmann, and Stefan Kollmannsberger \\
\textit{Chair of Computational Modeling and Simulation, Technical University of Munich, Germany}}
\date{}

\maketitle

\begin{abstract}
    \noindent
    Ductile damage models and cohesive laws incorporate the material plasticity entailing the growth of irrecoverable deformations even after complete failure. This unrealistic growth remains concealed until the unilateral effects arising from the crack closure emerge. We address this issue by proposing a new strategy to cope with the entire process of failure, from the very inception in the form of diffuse damage to the final stage, i.e. the emergence of sharp cracks. To this end, we introduce a new strain field, termed \textit{discontinuity strain}, to the conventional additive strain decomposition to account for discontinuities in a continuous sense so that the standard principle of virtual work applies. We treat this strain field similar to a strong discontinuity, yet without introducing new kinematic variables and nonlinear boundary conditions. In this paper, we demonstrate the effectiveness of this new strategy at a simple ductile damage constitutive model. The model uses a scalar damage index to control the degradation process. The discontinuity strain field is injected into the strain decomposition if this damage index exceeds a certain threshold. The threshold corresponds to the limit at which the induced imperfections merge and form a discrete crack. With three-point bending tests under pure mode I and mixed-mode conditions, we demonstrate that this augmentation does not show the early crack closure artifact which is wrongly predicted by plastic damage formulations at load reversal. We also use the concrete damaged plasticity model provided in Abaqus commercial finite element program for our comparison. Lastly, a high-intensity low-cycle fatigue test demonstrates the unilateral effects resulting from the complete closure of the induced crack.
\end{abstract}

\smallskip

\section{Introduction} \label{sec:introduction}
The last decades have witnessed extensive studies on computational failure mechanics. With ever growing computational resources at our disposal, various advanced numerical tools representing different aspects of failure problems have emerged, enabling us to predict complex failure scenarios. The complexity arises from the variety of stages involved, comprising the presence of diffuse imperfections, localization of intense deformations, and the formation of distinct sharp cracks. There is a spectrum of options to cope with the failure process, ranging from purely continuous-based approaches to purely discontinuous ones \cite{cervera2022comparative}. On the one hand, continuous approaches span the damaged area over a finite width of the medium from the emergence of induced defects, up to the formation of discontinuity surfaces. Therefore, they recast discrete cracks into narrow bands of highly localized deformation with (almost) zero stiffness. Continuum damage models \cite{bui2022dynamic, deng2023adaptive, daneshyar2023fracture} and phase field methods \cite{Hennig2022, kalina2023overview, dammass2023phase, loehnert2023enriched, haghighat2023efficient, schapira2023performance} belong to this group. On the other hand, discontinuous approaches represent the fracture process zone in the form of displacement jumps, no matter if there truly exists an abrupt change in the displacement field or not. Some prominent tools belonging to this group are the embedded discontinuity model \cite{panto2022two, bach2022embedded, nikolic2022discrete} and the extended finite element method \cite{benvenuti2021mesh, schmidt2023extended, bento2023recovery}.

Neither of the aforementioned strategies is self-sufficient in dealing with the entire fracture process. Continuous approaches smear sharp cracks over a finite width of the medium, while discontinuous approaches sample the damaged region on the fictitious crack faces. Both represent the two extreme ends of the spectrum of physical effects well and their strengths and deficiencies complement each other to some extent. Therefore, numerous studies have been devoted to leveraging both approaches by establishing a transition from distributed damage to strong discontinuities. Simone et al. \cite{simone2003continuous} alleviated the unrealistic damage growth resembled by the class of continuous models through combining a regularised continuum framework with a partition of unity finite element method. Introducing the thick level set approach, Mo{\"e}s et al. \cite{moes2011level} allowed a straightforward transition from damaged zone to complete fracture once the material is totally damaged. Han et al. \cite{han2023transition} proposed a continuous-discontinuous framework considering large deformation kinematics to recast diffuse degraded topologies into sharp crack paths. Dynamic crack propagation induced by gradient-enhanced damage growth was presented by Sun and L{\"o}hnert \cite{sun20213d}. They coupled transient thermal and dynamic mechanical fields and, used the extended finite element method in three-dimensional settings for representing discrete cracks. Pandey et al. \cite{pandey2023new} introduced a hybrid methodology by incorporating the extended finite element method and continuum damage mechanics to represent creep-fatigue crack propagation. A continuous-discontinuous strategy within which a thin layer represents the discontinuity emerging due to localized failure is presented by Puccia et al. \cite{puccia2023finite}. This layer behaves similar to the bulk material so that no additional constitutive model associated with cohesive-like approaches is required. Regarding the conservation of energy during the transition, some continuous-discontinuous approaches inject the discontinuity at the final stage of the fracture process when the material is almost fully degraded to avoid spurious energy release. The combined model of Seabra et al. \cite{seabra2013damage} for damage-driven crack propagation in ductile metals, the fracture-based continuous-discontinuous approach of Sarkar et al. \cite{sarkar2021simplified}, and the enriched continuum model of Negi and Kumar \cite{negi2022continuous} fall into this category. Others use the concept of energy equivalence to define an intermediate state so that the transition can be triggered at any stage in a smooth manner. The hybrid model of Cuvilliez et al. \cite{cuvilliez2012finite}, the continuum to discontinuum transition strategy of Roth et al. \cite{roth2015combined}, and the thermodynamically consistent model of Wang and Waisman \cite{wang2016diffuse} are some examples of this type.
    
\begin{figure*} [ht]
    \centering
    \includegraphics[scale=1.0]{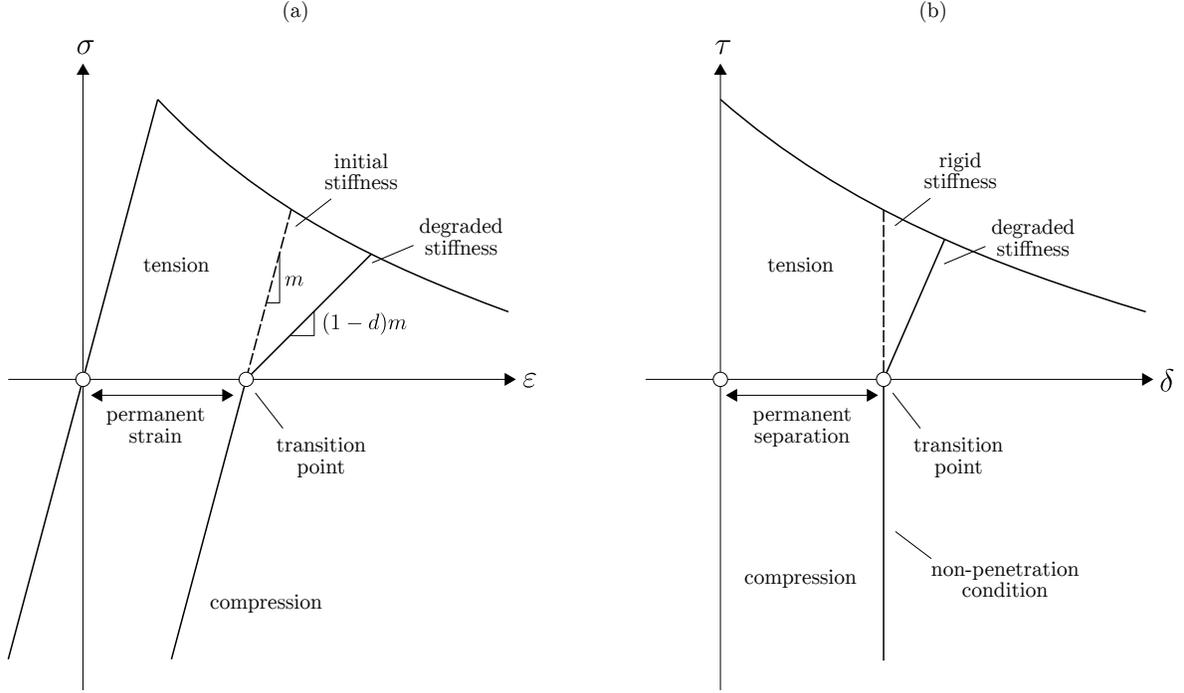}
    \caption{Constitutive response of generic ductile damage models and cohesive laws: (a) stress-strain response, and (b) traction-separation law.}
    \label{fig:unilateral}
\end{figure*}

\begin{figure*} [ht]
    \centering
    \includegraphics[scale=1.0]{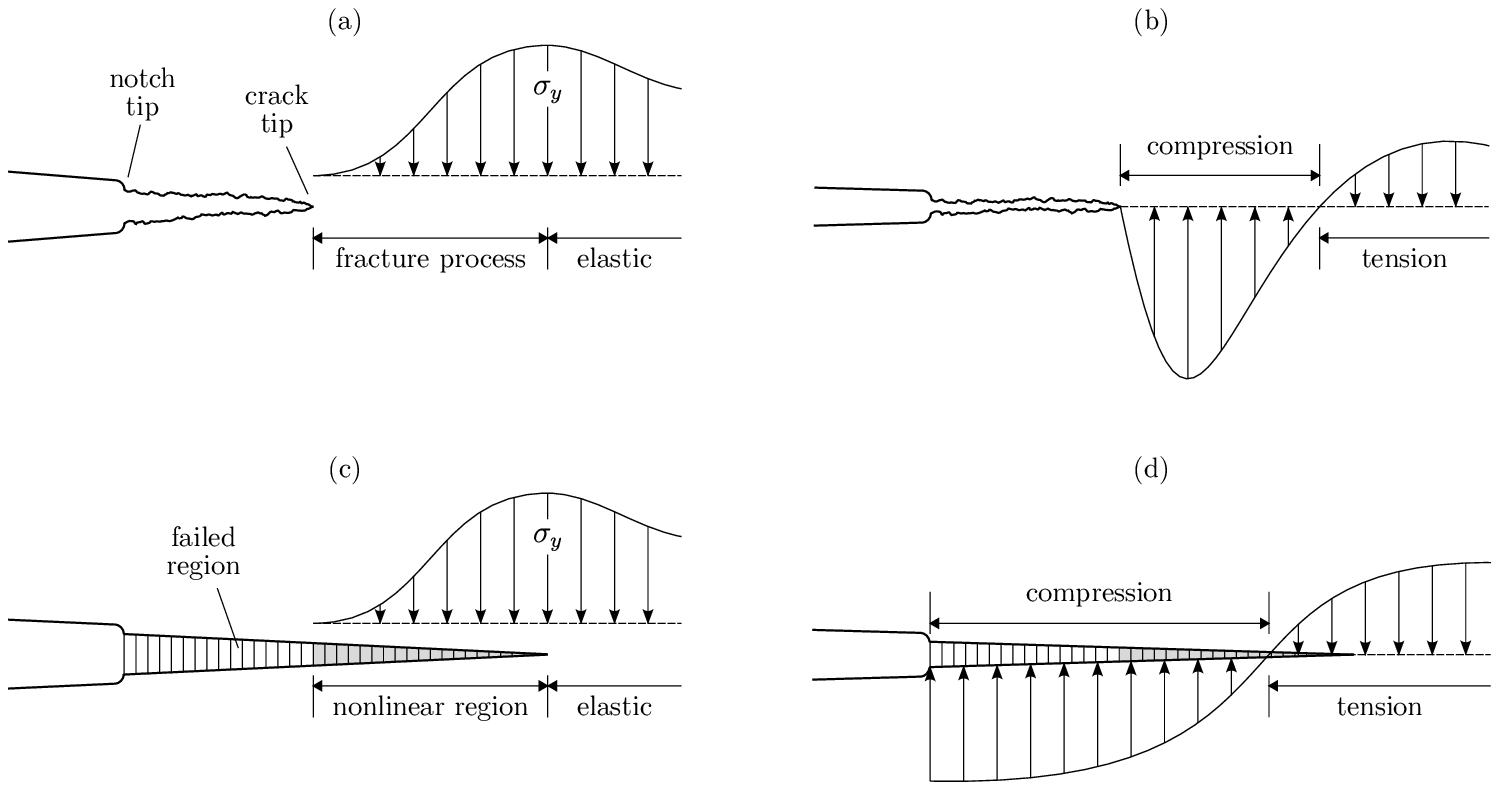}
    \caption{State of a crack under loaded and relaxed conditions: (a) realistic loaded state, (b) realistic relaxed state, (c) numerical loaded state, and (d) numerical relaxed state.}
    \label{fig:fracture}
\end{figure*}

All of the above and numerous other can accurately address the failure process over moderately complex loading scenarios. Yet they are not necessarily well-equipped to deal with complex loading paths. The most overlooked aspect of the fracture process in the strategies mentioned is the material response during load reversal. Elegant constitutive models and cohesive laws incorporate plasticity by which irreversible strains and displacement jumps are coupled with reversible ones \cite{cervera2015conformity}. As a result, all continuous, discontinuous or hybrid models exhibit excessive unrealistic permanent deformations by construction which are accumulated within the smear or on the discrete fracture process zone. Figure \ref{fig:unilateral} depicts a generic stress-strain response and traction-separation law of such models. The crack closure effects lead to stiffness recovery upon transition from tension to compression. This unilateral behavior is essential in general loading scenarios, such as cyclic loading \cite{hartmann1998numerical, cervera2018cracking, alaimo2021numerical, zhang2023novel}, fatigue tests \cite{wachter2020monotonic, oneschkow2022compressive, schroder2022phase, pise2023phenomenological}, and seismic modeling \cite{daneshyar2021wave, coronelli2023flat, schiavoni2023advanced}, that entail complex strain paths. Obeying the responses shown, fully degraded regions develop excessive permanent deformations that shift the transition point to the right of the curve. As a result, the unilateral effects may get triggered in a faulty state and lead to artificial stiffening. This is elaborated further in Figure \ref{fig:fracture}, which shows the normal stress distribution along the crack axis. According to the figure, the fracture process zone nucleated in front of the crack tip spans over the domain that experiences a nonlinear response. As a consequence of balancing the internal forces, part of this zone experiences compressive stresses upon external load removal (see Figure \ref{fig:fracture}~(b)). However, the crack faces spanning from the notch tip to the crack tips remain traction-free in this relaxed state. This is not the case for the relaxed numerical model as the fully degraded section of the crack develops compressive stresses that affect the global response of the numerical model. This artificial stiffening remains untraced when a constantly increasing load is applied and can only be detected in scenarios that involve unloading. A remedy is to limit the growth of permanent deformations to an ultimate state corresponding to the limit at which discrete crack faces emerge. This treatment is completely impractical in the continuous models, yet can be applied in the discontinuous ones by permanently detaching the opposing crack faces and removing the stress continuity condition. In this case, resembling the unilateral effects still invokes the use of computational contact mechanics, which renders the model even more complex.

By contrast, the paper at hand introduces the discontinuity in the strain field to mimic discrete crack faces. As a result, no nonlinear boundary conditions or kinematic variables are involved, preserving the standard form of the virtual work principle. To this end, we organize the body of this text as follows. After this introduction, the additive strain decomposition including the discontinuity strain field is introduced in Section \ref{sec:theory}, followed by the underlying theory of a ductile damage model involving the unilateral effects. It is worth mentioning that a simple damage model is chosen to avoid unnecessary complications arising from the constitutive modeling. However, the discontinuity strain field can be incorporated into other material models regardless of the hypothesis involved. Section \ref{sec:examples} is devoted to numerical examples. A three-point bending specimen under pure opening mode is analyzed first. Then, a similar test is conducted under mixed-mode conditions. The complete closure of the crack is also resembled by applying a high-intensity low-cycle load on a double-edge notched specimen before conclusions drawn in Section \ref{sec:conclusion} close the paper.

\section{Theory} \label{sec:theory}
\begin{figure*} [ht]
    \centering
    \includegraphics[scale=1.0]{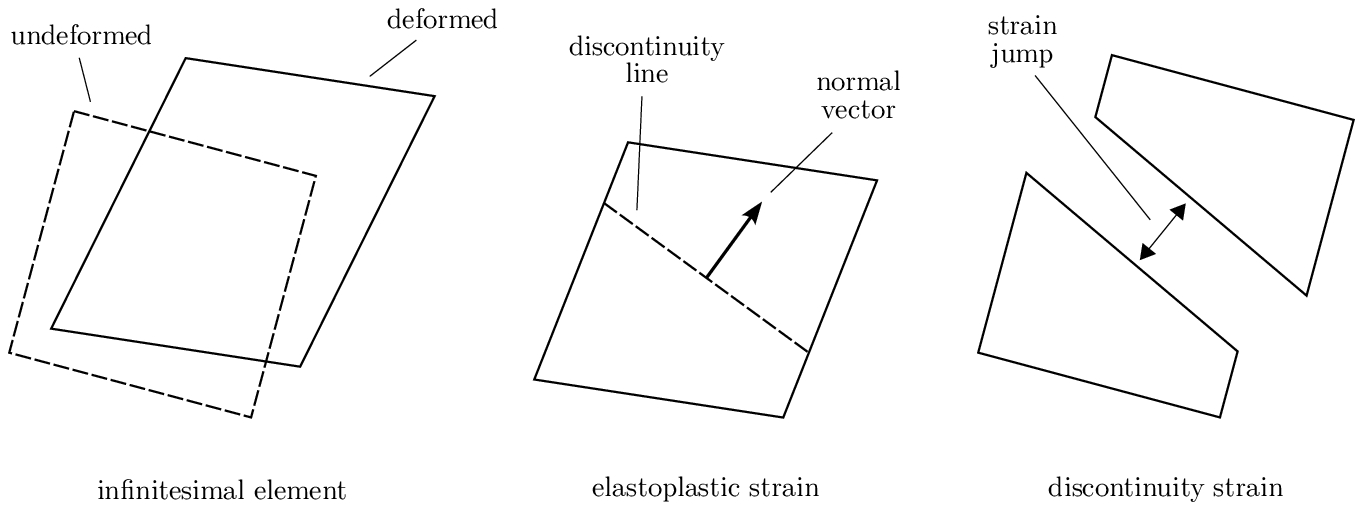}
    \caption{An abstraction of the additive strain split including the discontinuity strain field.}
    \label{fig:infinitesimal}
\end{figure*}

Material inelasticity involves both reversible and non-reversible changes of shape in the consequence of applied loads. The classical models of infinitesimal plasticity characterize these two by means of an additive split with which the total strain is decomposed into an elastic and a plastic part. We augment this additive strain decomposition by introducing the discontinuity strain field $\bm\varepsilon^d$ so that the total strain tensor reads
\begin{equation}
    \bm\varepsilon = \bm\varepsilon^e + \bm\varepsilon^p + \bm\varepsilon^d,
\end{equation}
where $\bm\varepsilon^e$ and $\bm\varepsilon^p$ are the conventional elastic and plastic parts, respectively. Here, the strain field $\bm\varepsilon^d$ represents the induced crack due to the excessive damage growth, but in general, it can mimic any kind of discontinuity. In contrast to the plastic strain $\bm\varepsilon^p$, the discontinuity strain field is reversible, allowing it to resemble the opening and closing of the hypothetical crack. Figure \ref{fig:infinitesimal} shows an abstraction of this additive decomposition. By activating the discontinuity strain field at a certain damage threshold, which corresponds to the limit of a sharp crack forming due to the complete detachment of material, further plastic straining is prevented. As a consequence, crack closure can occur at this detachment limit in the case of load reversal. Here, we incorporate this strain split in a simple isotropic damage model to keep the constitutive modeling minimal. However, it is worth mentioning once again that the discontinuity strain field can be used in any material model and also for representing any kind of discontinuity.

Following the concept of effective quantities originally introduced by Kachanov \cite{kachanov1958}, we characterize the material elastoplastic regime within a fictitious intact configuration at which the mechanical stress, known as the effective stress $\tilde{\bm\sigma}$, is the measure of average force acting on the \textit{undamaged} cross-sectional area of a body. Hence, whether the infinitesimal volume element of material is influenced by damage or not, the following general constitutive law of isotropic material applies
\begin{equation}
    \tilde{\bm\sigma} = \tilde{\mathbf{D}}:\bm\varepsilon^e,
\end{equation}
or, equivalently,
\begin{equation}
    \tilde{\bm\sigma} = \tilde{\mathbf{D}}:(\bm\varepsilon - \bm\varepsilon^p - \bm\varepsilon^d),
\end{equation}
where $\tilde{\bf{D}}$ is the tangent operator of intact material. Note that we preserve this naming convention throughout the text so that the tilde accent refers to the effective configuration.

According to the work of Feenstra and de Borst \cite{feenstra1995plasticity}, the yield locus is given in the effective configuration by means of the Rankine maximum principal stress criterion through
\begin{equation} \label{eq:yield}
    f(\tilde{\bm\sigma}) = \tilde{\sigma}_{max} - \sigma_y,
\end{equation}
where $\sigma_y$ is the yield strength. Hence, the admissibility of the stress state must be preserved by enforcing it to remain inside or on this locus. To this end, the evolution of the plastic strain is given by the associated flow rule
\begin{equation}
    \dot{\bm\varepsilon}^p = \dot\gamma\partial_{\tilde{\bm\sigma}}f,
\end{equation}
wherein $\dot\gamma$ is a Lagrange multiplier. The pair $f$ and $\dot\gamma$ is subjected to the Kuhn--Tucker optimality conditions
\begin{equation}
    f \le 0,\qquad \dot\gamma \ge 0,\qquad \dot\gamma f = 0 ,
\end{equation}
which enforce the stress state admissibility determining whether yielding occurs or not. Now by defining
\begin{equation}
    \bm\sigma = (1-d)\tilde{\bm\sigma},
\end{equation}
the stress tensor in the effective configuration can be mapped to its macroscopically observed counterpart, the Cauchy stress tensor $\bm\sigma$. In the above mapping, the damage index $d$ represents the density of imperfections reducing the load-bearing area of the infinitesimal volume element of the material. However, in order to establish the unilateral conditions, the above relation must be modified such that the damage index $d$ only applies if the stress state is in tension. To this end, the stress split
\begin{equation} \label{eq:stress-split}
    \tilde{\bm\sigma} = \tilde{\bm\sigma}_t + \tilde{\bm\sigma}_c
\end{equation}
is defined wherein the subscripts $t$ and $c$ denote the tensile and compressive parts, respectively \cite{de2011computational}. The tensile part can be given by \cite{ortiz1985constitutive}
\begin{equation}
    \tilde{\bm\sigma}_t = \sum_{i=1}^3{\langle \tilde{\sigma}_i\rangle\bm{e}_i\otimes\bm{e}_i},
\end{equation}
wherein
\begin{equation}
    \langle x \rangle = \left\{ 
    {\begin{array}{*{20}{c}}
    x&{x  >  0}\\
    0&{x \le 0}
    \end{array}} \right.
\end{equation}
is the Macaulay bracket, and $\tilde{\sigma}_i$ and $\bm{e}_i$ are the $i^\text{th}$ principal value and principal direction, respectively. Employing the additive split \eqref{eq:stress-split}, the compressive part reads
\begin{equation}
    \tilde{\bm\sigma}_c = \tilde{\bm\sigma} - \sum_{i=1}^3{\langle \tilde{\sigma}_i\rangle\bm{e}_i\otimes\bm{e}_i}.
\end{equation}
With the tensile and compressive parts at hand, the stress mapping including the unilateral effects is given by
\begin{equation}
    \bm\sigma = (1-d)\tilde{\bm\sigma}_t + \tilde{\bm\sigma}_c.
\end{equation}
As a result, due to the micro-cracks closure upon transition from tension to compression, the initial stiffness of the material is recovered.

The damage evolution is closely linked to the growth of plastic strain \cite{daneshyar2020fe}. However, depending on the material being studied, its growth may start as the plastic strain is mobilized or when the yielding process reaches a certain stage. Inspired by the work of Grassl and Jir\'{a}sek \cite{grassl2006plastic}, we use the exponential law
\begin{equation} \label{eq:exponential1}
    d = 1-\exp(-a\bar\varepsilon^p),
\end{equation}
where $a$ is considered as a material constant identifying the magnitude of damage growth, and
\begin{equation} \label{eq:accumulated}
    \bar\varepsilon^p = \int_t{\dot\gamma dt}
\end{equation}
is the accumulated equivalent plastic strain. It is worth mentioning that, for a model aiming at realistic constitutive modeling, a more complicated relation can be opted for. Now, at the onset of cracking we have
\begin{equation}
    d_{cr} = 1-\exp(-a\bar\varepsilon^p_{cr}),
\end{equation}
where $d_{cr}$ is the critical damage denoting the threshold at which the induced imperfections merge and form a sharp crack. Utilizing the above expression, we can define the critical accumulated equivalent plastic strain at the inception of cracking as
\begin{equation}
    \bar\varepsilon^p_{cr} = -\frac{1}{a}\ln{(1-d_{cr})}.
\end{equation}
Thus, the discontinuity strain field mobilizes if the accumulated equivalent plastic strain reaches the above limit. Referring to \eqref{eq:accumulated}, for two arbitrary successive (pseudo-)time steps $t$ and $t+dt$ we have
\begin{equation}
    \bar\varepsilon^p_{t+dt} = \bar\varepsilon^p_{t} + \dot\gamma.
\end{equation}
However, the accumulated equivalent plastic strain cannot exceed the limit $\bar\varepsilon^p_{cr}$. Hence, if the standard return mapping causes an overshoot, there exists an intermediate state between these two successive steps, identified by $t + \alpha dt$, at which
\begin{equation} \label{eq:intermediate}
    \bar\varepsilon^p_{t+\alpha dt} = \bar\varepsilon^p_{cr},
\end{equation}
and the maximum principal stress hits the yield locus at the same time. Defining $\lambda$ that denotes the evolution of $\bar\varepsilon^p$ from the time step $t$ to the time step $t + \alpha dt$, we arrive at
\begin{equation}
    \bar\varepsilon^p_{t+\alpha dt} = \bar\varepsilon^p_t + \lambda.
\end{equation}
By plugging the magnitude of $\bar\varepsilon^p_{t+\alpha dt}$ from \eqref{eq:intermediate} into the above expression, $\lambda$ is obtained as
\begin{equation}
    \lambda = \bar\varepsilon^p_{cr} - \bar\varepsilon^p_t.
\end{equation}
As a result, the growth of the equivalent plastic strain is known beforehand and the evolution of the plastic strain from $t$ to $t + \alpha dt$ can be given explicitly by
\begin{equation}
    \dot{\bm\varepsilon}^p = \lambda\partial_{\tilde{\bm\sigma}}f.
\end{equation}
Hence, instead of solving the standard nonlinear return mapping, we must find $\alpha$ such that
\begin{equation}
    f(\tilde{\bm\sigma}_{t+\alpha dt}) = 0,
\end{equation}
wherein
\begin{equation}
    \tilde{\bm\sigma}_{t+\alpha dt} = \tilde{\bm\sigma}_{t} + \tilde{\mathbf{D}}:(\alpha\dot{\bm\varepsilon} - \lambda\partial_{\tilde{\bm\sigma}}f).
\end{equation}
From the intermediate state $t+\alpha dt$ onward, further strains are accumulated on the discontinuity strain field. As a result, the discontinuity strain $\bm\varepsilon^d$ emerges at $t+\alpha dt$ and receives the remaining part of the total strain rate during $t+\alpha dt$ to $t+dt$ so that
\begin{equation}
    \bm\varepsilon^d_{t+dt} = (1-\alpha)\dot{\bm\varepsilon}.
\end{equation}
In addition, to comply with the Rankine maximum principal stress criterion, the discontinuity plane is set to the plane of maximum tensile stress at the intermediate state $t+\alpha dt$. The normal vector to this plane, denoted by $\bm{n}$, will be used in the subsequent stages of the analysis to check whether the crack is open or not. Crack closure is detected if
\begin{equation}
    \bm{n}^\intercal\bm\varepsilon^d \bm{n} < 0,
\end{equation}
meaning that the opposing crack faces penetrate each other. Once again, we must find an intermediate step, denoted by $t+\beta dt$, at which the opposing crack faces meet. By writing the incremental evolution of the discontinuity strain field from $t$ to $t+\beta dt$, we arrive at
\begin{equation}
    \bm\varepsilon^d_{t+\beta dt} = \bm\varepsilon^d_{t} + \beta\dot{\bm\varepsilon}.
\end{equation}
Solving
\begin{equation}
    \bm{n}^\intercal \bm\varepsilon^d_{t} \bm{n} + \beta\bm{n}^\intercal \dot{\bm\varepsilon} \bm{n} = 0
\end{equation}
for $\beta$, the exact time of crack closure is obtained. After crack closure, the discontinuity strain field vanishes and the standard elastoplasticity applies until the inception of another crack opening regime. It should be noted that, by employing this formulation, the maximum damage that can be reached is limited to its critical value since no plastic straining occurs afterward. As a result, the degradation process freezes once the damage index reaches its upper limit, which is only acceptable if $d_{cr}$ is chosen extremely close to unity. Otherwise the material preserves its load-bearing capacity and never degrades completely. To allow further damage growth, the exponential function \eqref{eq:exponential1} is redefined as
\begin{equation} \label{eq:exponential2}
    d = 1-\exp(-a\bar\varepsilon^p)\exp(-b\bar\varepsilon^c),
\end{equation}
where $b$ is a dimensionless constant used to adjust the damage growth after cracking, and
\begin{equation}
    \bar\varepsilon^d = \max_{\tau \le t}(\bm{n}^\intercal\bm\varepsilon^d \bm{n}),
\end{equation}
is the maximum strain jump that is experienced during the loading history. Needless to say that a proper strategy is to establish a relation between the damage growth function \eqref{eq:exponential2} and the fracture energy of material so that the constants $a$ and $b$ get linked to a physical property. In addition, it can provide the objectivity of global responses with respect to numerical discretization. However, since this paper is not aimed at the constitutive modeling of material, we will pursue this target in a separate paper in the future.

\section{Numerical results} \label{sec:examples}
This section is devoted to showing the contribution of the discontinuity strain field in the global responses of specimens undergoing the failure process. To this end, three laboratory tests are chosen, including a pure mode I three-point bending test, a mixed-mode test, and a double-edge notched specimen subjected to cyclic loading. The standard finite element method is used to discretize the domain of problems. Note that the solutions are not unique since the softening response during the failure process induces a local material instability that leads to the ill-posedness of the boundary value problem. This can be cured by injecting a measure of length, known as the material length scale, representing the width of the fracture process zone through non-local or gradient-enhanced formulations or by using the concept of fracture energy equivalence \cite{daneshyar2023scaled}. In the paper at hand, this ill-posedness is left untreated to focus only on the introduced discontinuity strain field and keep the message of the paper as clear as possible. As a result, due to the unregularized nature of the field equations, the damage constants $a$ and $b$ are calibrated in accordance with the finite element meshes. This mesh-dependency vanishes once a material length scale is introduced.

We analyze each test using different assumptions of the critical damage $d_{cr}$ to show its effects on the unloading branches. To consider the cases in which the conventional strain decomposition is used, the critical damage $d_{cr}$ is set to unity. As a result, since $d$ is defined by means of an exponential function that never reaches this ultimate value (see the damage growth function in \eqref{eq:exponential2}), the discontinuity strain field is simply neglected in the simulations. In addition, to further illustrate the shortcoming of the conventional formulations, the concrete damaged plasticity model of Abaqus finite element program is also used to simulate the tests. This model is capable of reproducing the unilateral effects, which is the key ingredient for revealing the artificial stiffening upon load reversal. Note that one-dimensional stress-strain and damage-strain curves are required to define the post-linear behavior in the Abaqus model. These curves are provided for each example using the calibrated values of $E$, $\sigma_y$, and $a$ in conjunction with the one-dimensional Hooke's law and the exponential damage growth function in \eqref{eq:exponential1}. Other parameters of the model are set to their default values.

\subsection{Opening-mode test}
\begin{figure*}
    \centering
    \includegraphics[scale=1.0]{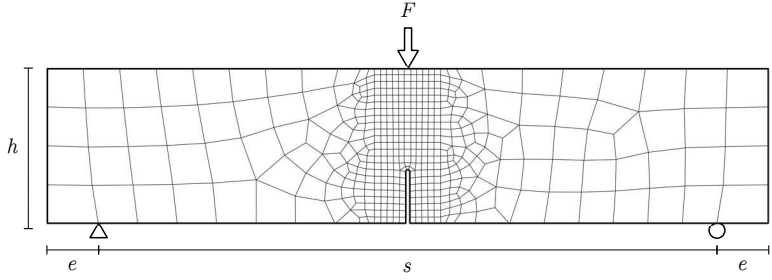}
    \caption{Geometry, boundary conditions, and finite element mesh of the opening-mode test.}
    \label{fig:pure_mesh}
\end{figure*}

This numerical example is a replication of the laboratory test of Perdikaris and Romeo \cite{perdikaris1995size}. A series of concrete beams with similar dimensions, yet notched at different offsets were tested. We use the mid-span notched beam for testing the model under a pure mode I loading. Figure \ref{fig:pure_mesh} shows the finite element model of the beam on the test setup. The dimensions $s$, $h$, and $e$ are 304.8, 76.2, and 25.4 millimeters, respectively. In addition, the notch depth is one-third of the height of the beam. The test is simulated under plane stress conditions with a thickness of 28.6 millimeters. The material properties of the beam are reported in Table \ref{tab:pure}. Four cases of $d_{cr}$ are considered, including $1.00$, $0.85$, $0.60$, and $0.45$. The former, $d_{cr}=1.0$, denotes the condition in which the discontinuity strain field is disregarded. Decreasing the value of $d_{cr}$, the contribution of the discontinuity strain field becomes more pronounced. The test is also modeled by means of the concrete damaged plasticity model of Abaqus.

\begin{table}[h]
    \centering
    \caption{Parameters of the opening-mode test.} \label{tab:pure}
    \smallskip
    \begin{tabular}{@{}ccc@{}}
    \hline
    parameter   & value & unit  \\
    \hline
    $E$         & 28    & GPa   \\
    $\nu$       & 0.2   & ---   \\
    $\sigma_y$  & 3.8   & MPa   \\
    $a$         & 80    & ---   \\
    $b$         & 70    & ---   \\
    \hline
    \end{tabular}
\end{table}

\begin{figure*}
    \centering
    \includegraphics[scale=1.0]{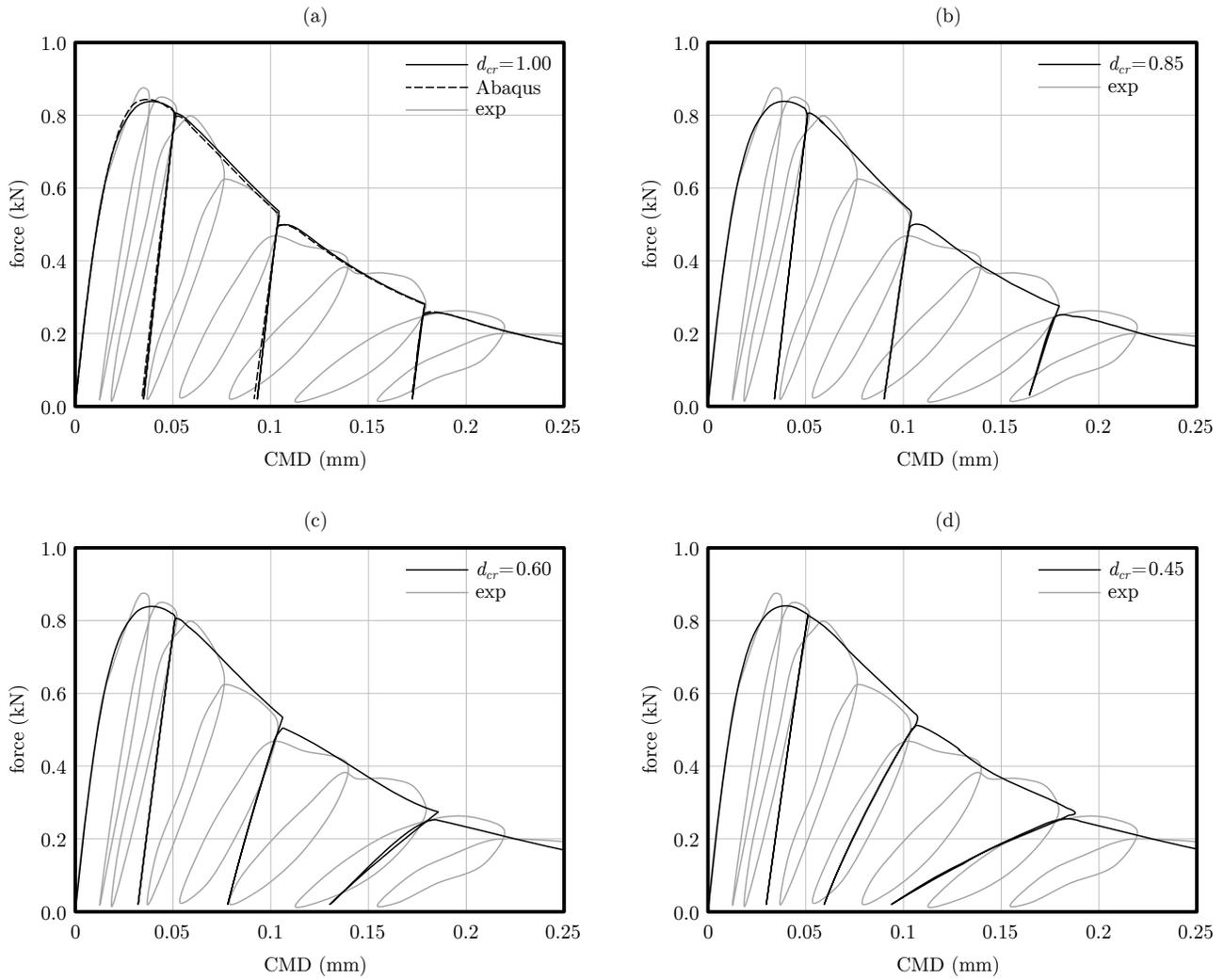}
    \caption{Comparison of the global responses with the experimental data for the opening-mode test: (a) Abaqus model and $d_{cr}=1.00$, (b) $d_{cr}=0.85$, (c) $d_{cr}=0.60$, and (d) $d_{cr}=0.45$.}
    \label{fig:pure_curves}
\end{figure*}

\begin{figure*}
    \centering
    \includegraphics[scale=1.0]{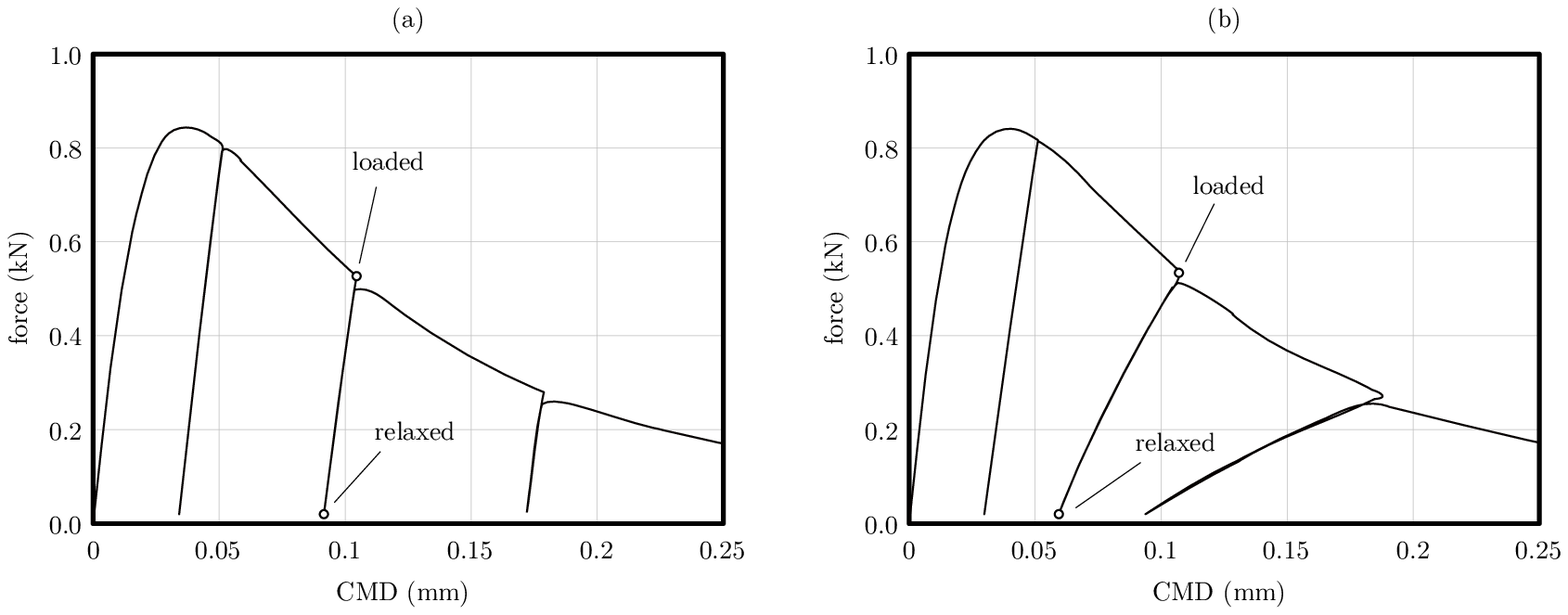}
    \caption{Loaded and relaxed states at the beginning and the end of the second unloading phase for the opening-mode test: (a) Abaqus model, and (b) $d_{cr}=0.45$.}
    \label{fig:pure_stages}
\end{figure*}

\begin{figure*}
    \centering
    \includegraphics[scale=1.0]{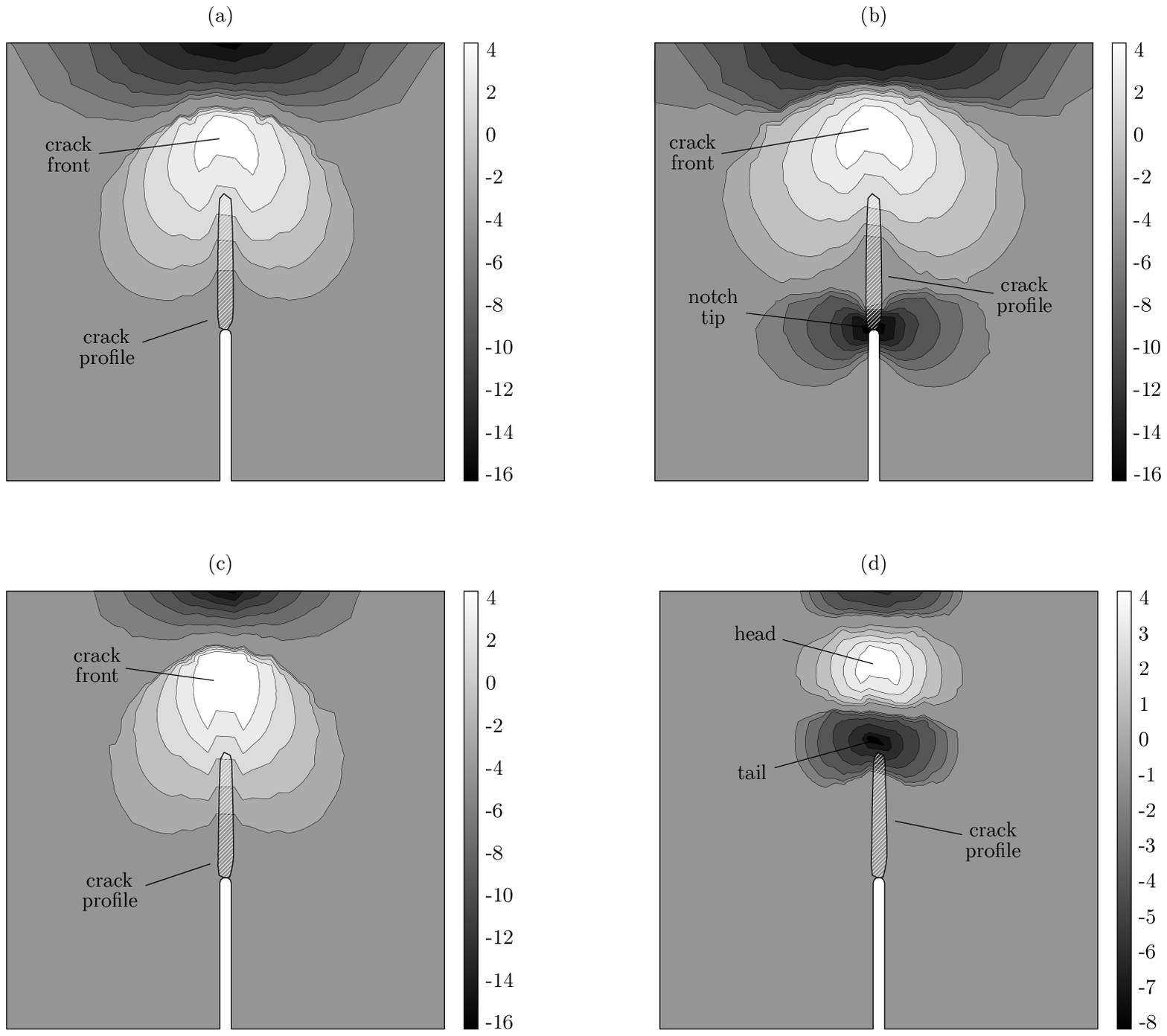}
    \caption{Stress contours (MPa) at the loaded and relaxed states for the opening-mode test: (a) loaded state in the Abaqus model, (b) relaxed state in the Abaqus model, (c) loaded state in the $d_{cr}=0.45$ case, and (d) relaxed state in the $d_{cr}=0.45$ case.}
    \label{fig:pure_contours}
\end{figure*}

The global responses of the numerical models are plotted against the experimental data in Figure \ref{fig:pure_curves}. The horizontal axis, i.e. the crack mouth displacement (CMD), represents the absolute relative displacement of the crack mouth. Figure \ref{fig:pure_curves}~(a) clearly shows that the conventional additive strain decomposition, which is the case in the model with $d_{cr}=1.0$ and the constitutive mode of Abaqus, causes the unloading branches of the curves to have a slope similar to the initial elastic branch. On the other hand, by decreasing the value of $d_{cr}$, which is accompanied by earlier activation of the discontinuity strain field, the unloading slopes gradually become less steep. The cause of this behavior is elaborated further by comparing the stress distributions before and after the load removal. To this end, consider the loaded and the corresponding relaxed states of the Abaqus model and the $d_{cr}=0.45$ case depicted in Figure \ref{fig:pure_stages}. The stress contours of these cases are shown in Figure \ref{fig:pure_contours}. Note that the absolute values of the principal stresses are used so that both tensile and compressive distributions are shown. Figure \ref{fig:pure_contours}~(a) depicts the stress distributions at the beginning of the unloading phase for the Abaqus model and Figure \ref{fig:pure_contours}~(c) for the case of including the discontinuity strain field with the critical damage $d_{cr}=0.45$. The domain occupied by the fully formed crack is almost stress-free in both cases and the crack front experiences tensile stress with a peak value identical to the yield strength. However, in the relaxed state of the Abaqus model (Figure \ref{fig:pure_contours}~(b)), the notch tip experiences a considerable amount of compressive stress. Hence, a transition from compression to tension exists by moving along the crack axis from the notch tip to the crack front. However, the material must be completely detached and remain stress-free along some part of this axis. This faulty behavior is caused due to the unrealistic growth of plastic strain within the fully degraded regions. This excessive plastic strain causes an early activation of the unilateral effects upon a slight reduction of the total strain, leading to an unrealistically early transition from tension to compression. On the other hand, according to Figure \ref{fig:pure_contours}~(d), by including the discontinuity stain field, some part of the fully damaged region remains stress-free after unloading, and the core of the compressive region is shifted from the notch tip to the middle of the crack. The damaged region located between the tensile and compressive cores can be interpreted as the active fracture process zone, and these two cores indicate the head and tail of this zone. Referring to the schematic stress distributions shown previously in Figure \ref{fig:fracture}, we can deduce that by considering the discontinuity strain filed, the fully detached section of the crack is reproduced within the model, which is not the case when the discontinuity strain field is not included.

\subsection{Mixed-mode test}
The second example is also selected from the work of Perdikaris and Romeo \cite{perdikaris1995size}. The test setup is similar to the previous one, with the exception that the notch is offset by 75.6 millimeters from the mid-span. The properties presented in Table \ref{tab:mixed} are used for the numerical model. Similar to the previous test, in addition to the Abaqus model, four cases of $d_{cr}=1.00$, $d_{cr}=0.85$, $d_{cr}=0.60$, and $d_{cr}=0.45$ are considered.

\begin{figure*}
    \centering
    \includegraphics[scale=1.0]{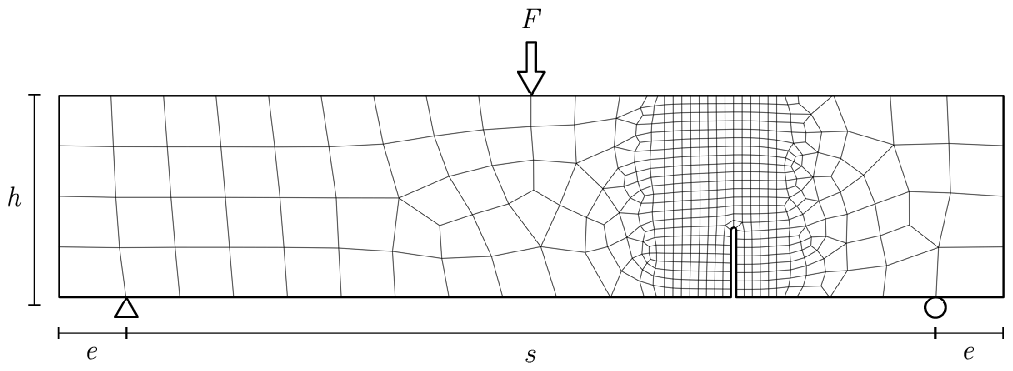}
    \caption{Geometry, boundary conditions, and finite element mesh of the mixed-mode test.}
    \label{fig:mixed_mesh}
\end{figure*}

\begin{table}[h]
    \centering
    \caption{Parameters of the mixed-mode test.} \label{tab:mixed}
    \smallskip
    \begin{tabular}{@{}ccc@{}}
    \hline
    parameter   & value & unit  \\
    \hline
    $E$         & 34    & GPa   \\
    $\nu$       & 0.2   & ---   \\
    $\sigma_y$  & 4.2   & MPa   \\
    $a$         & 110   & ---   \\
    $b$         & 70    & ---   \\
    \hline
    \end{tabular}
\end{table}

\begin{figure*}
    \centering
    \includegraphics[scale=1.0]{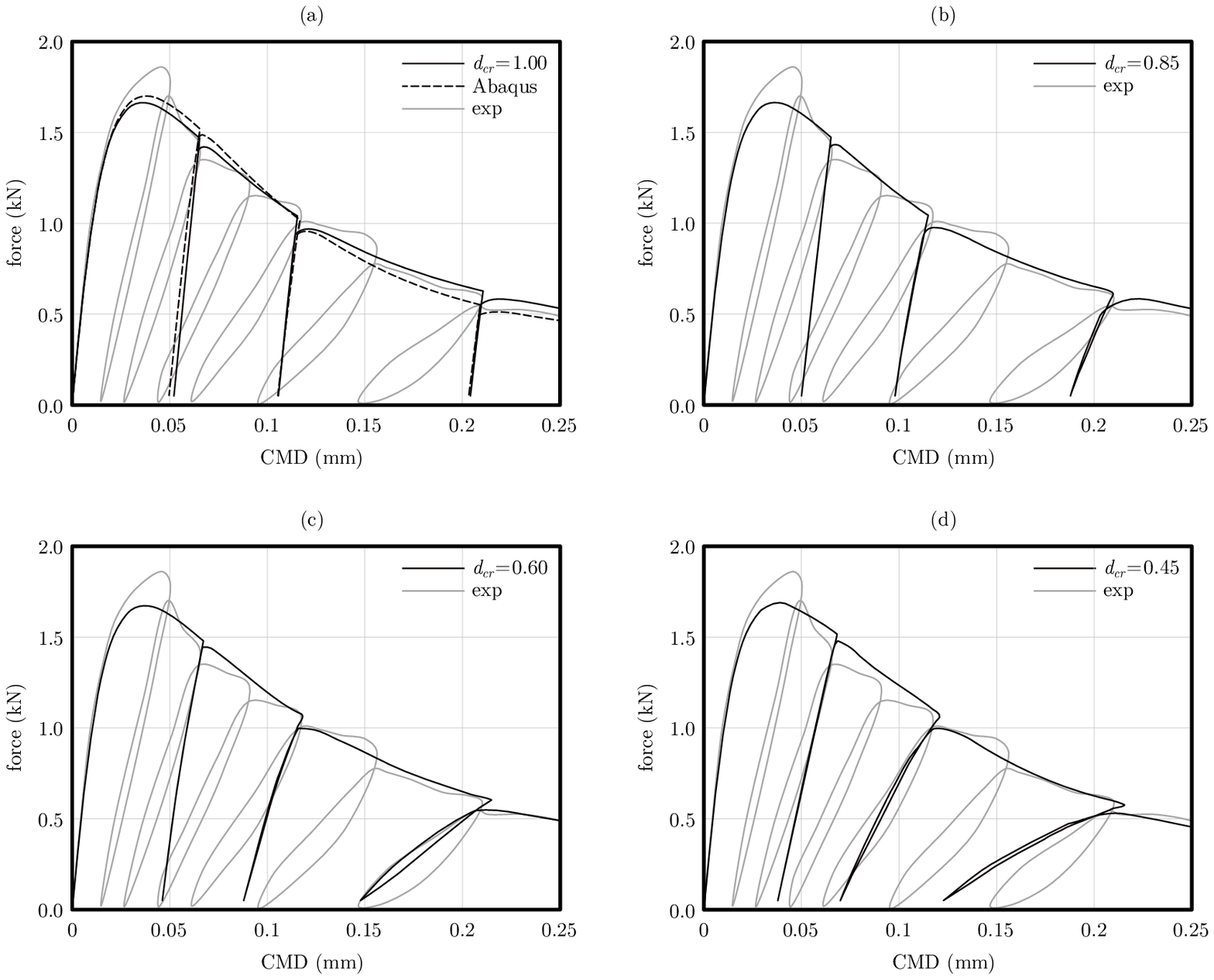}
    \caption{Comparison of the global responses with the experimental data for the mixed-mode test: (a) Abaqus model and $d_{cr}=1.00$, (b) $d_{cr}=0.85$, (c) $d_{cr}=0.60$, and (d) $d_{cr}=0.45$.}
    \label{fig:mixed_curves}
\end{figure*}

\begin{figure*}
    \centering
    \includegraphics[scale=1.0]{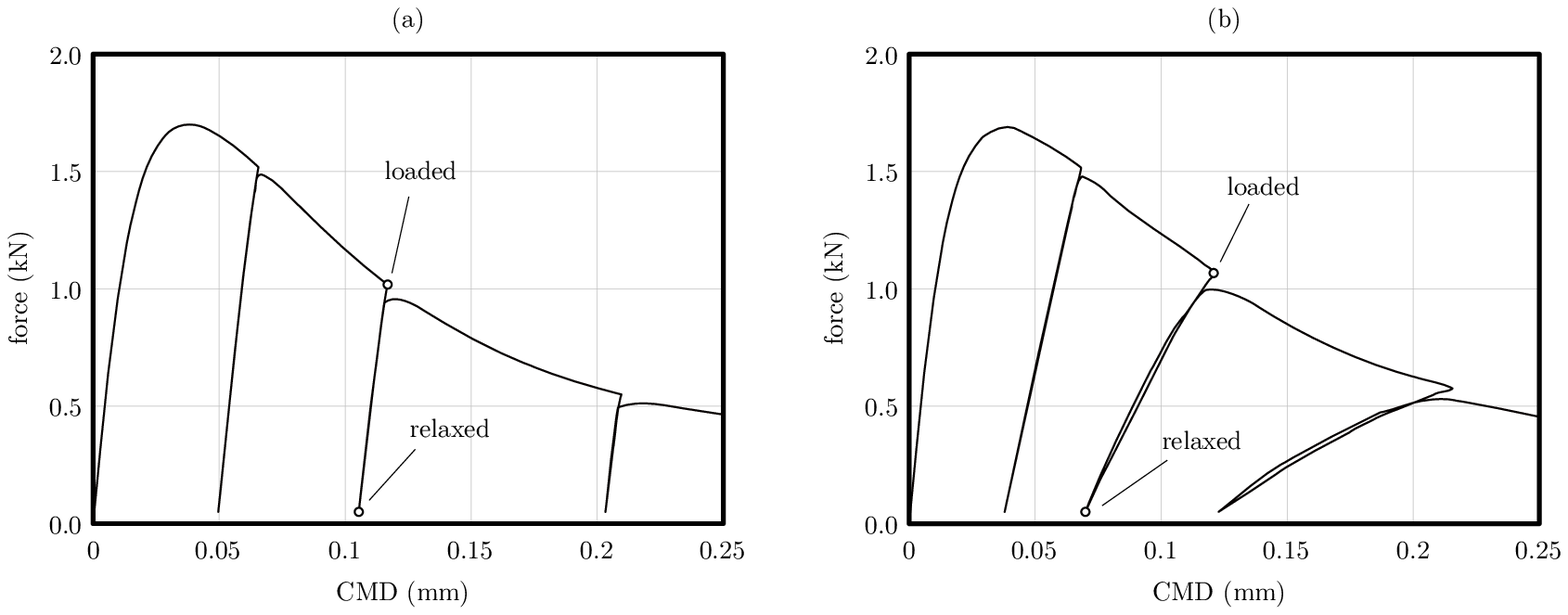}
    \caption{Loaded and relaxed states at the beginning and the end of the second unloading phase for the mixed-mode test: (a) Abaqus model, and (b) $d_{cr}=0.45$.}
    \label{fig:mixed_stages}
\end{figure*}

\begin{figure*}
    \centering
    \includegraphics[scale=1.0]{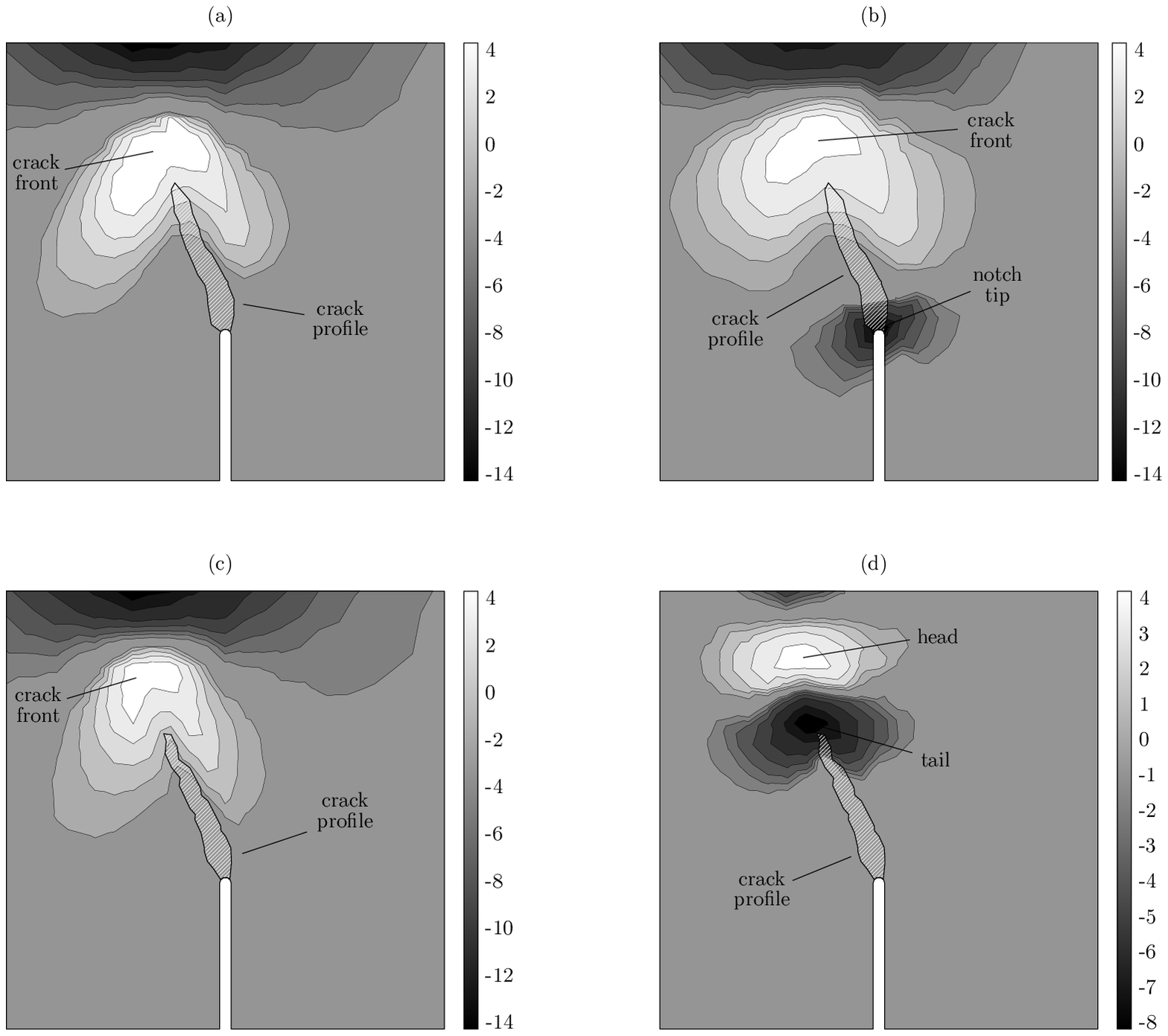}
    \caption{Stress contours (MPa) at the loaded and relaxed states for the mixed-mode test: (a) loaded state in the Abaqus model, (b) relaxed state in the Abaqus model, (c) loaded state in the $d_{cr}=0.45$ case, and (d) relaxed state in the $d_{cr}=0.45$ case.}
    \label{fig:mixed_contours}
\end{figure*}

The applied force versus the crack mouth displacement curves, plotted on the experimental data in Figure \ref{fig:mixed_curves}, conspicuously demonstrate the contribution of the discontinuity strain field. Referring to the stages marked in Figure \ref{fig:mixed_stages} for the Abaqus model and the $d_{cr}=0.45$ case, a deduction similar to that of the previous test regarding the stress contours shown in Figure \ref{fig:mixed_contours} can be drawn. Upon the external load removal, the fully degraded part of the crack develops compressive stress in the Abaqus model, while this part remains stress-free in the $d_{cr}=0.45$ case.

\subsection{Full-cycle test}
\begin{figure*}
    \centering
    \includegraphics[scale=1.0]{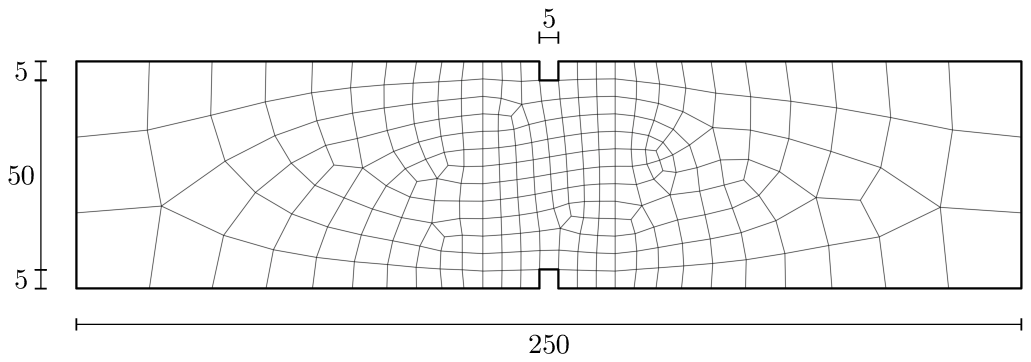}
    \caption{Geometry, boundary conditions, and finite element mesh of the full-cycle test.}
    \label{fig:fatigue_mesh}
\end{figure*}

The final example is dedicated to showing the crack closing and reopening behavior occurring in full-cycle tests. For this purpose, the high-intensity low-cycle fatigue test of Reinhardt \cite{reinhardt1984fracture}, conducted on double-edge notched specimens, is chosen. The test setup and finite element mesh of its numerical model are depicted in Figure \ref{fig:fatigue_mesh}. As shown in the figure, the length and width of the beam are 250 and 60 millimeters, respectively. The thickness of the beam is also 50 millimeters, included in the numerical model by means of the plane stress assumption.

The test is simulated by imposing prescribed opposing displacements on the left and right edges of the specimen. The recorded reaction forces are divided by the cross-section area of the specimen to represent the average stress. The average relative displacement of two vertical lines that are offset by 17.5 millimeters from the mid-section of the specimen is also recorded to generate the stress-deflection curves.

The contribution of the discontinuity strain field is assessed using four different values of $d_{cd}$, including $1.00$, $0.60$, $0.40$, and $0.20$. The concrete damaged plasticity model of Abaqus is also utilized to simulate the test. The material properties employed in the numerical model are presented in Table \ref{tab:double}.

\begin{table}[h]
    \centering
    \caption{Parameters of the full-cycle test.} \label{tab:double}
    \smallskip
    \begin{tabular}{@{}ccc@{}}
    \hline
    parameter   & value & unit  \\
    \hline
    $E$         & 25    & GPa   \\
    $\nu$       & 0.2   & ---   \\
    $\sigma_y$  & 3.2   & MPa   \\
    $a$         & 150   & ---   \\
    $b$         & 140   & ---   \\
    \hline
    \end{tabular}
\end{table}

\begin{figure*}
    \centering
    \includegraphics[scale=1.0]{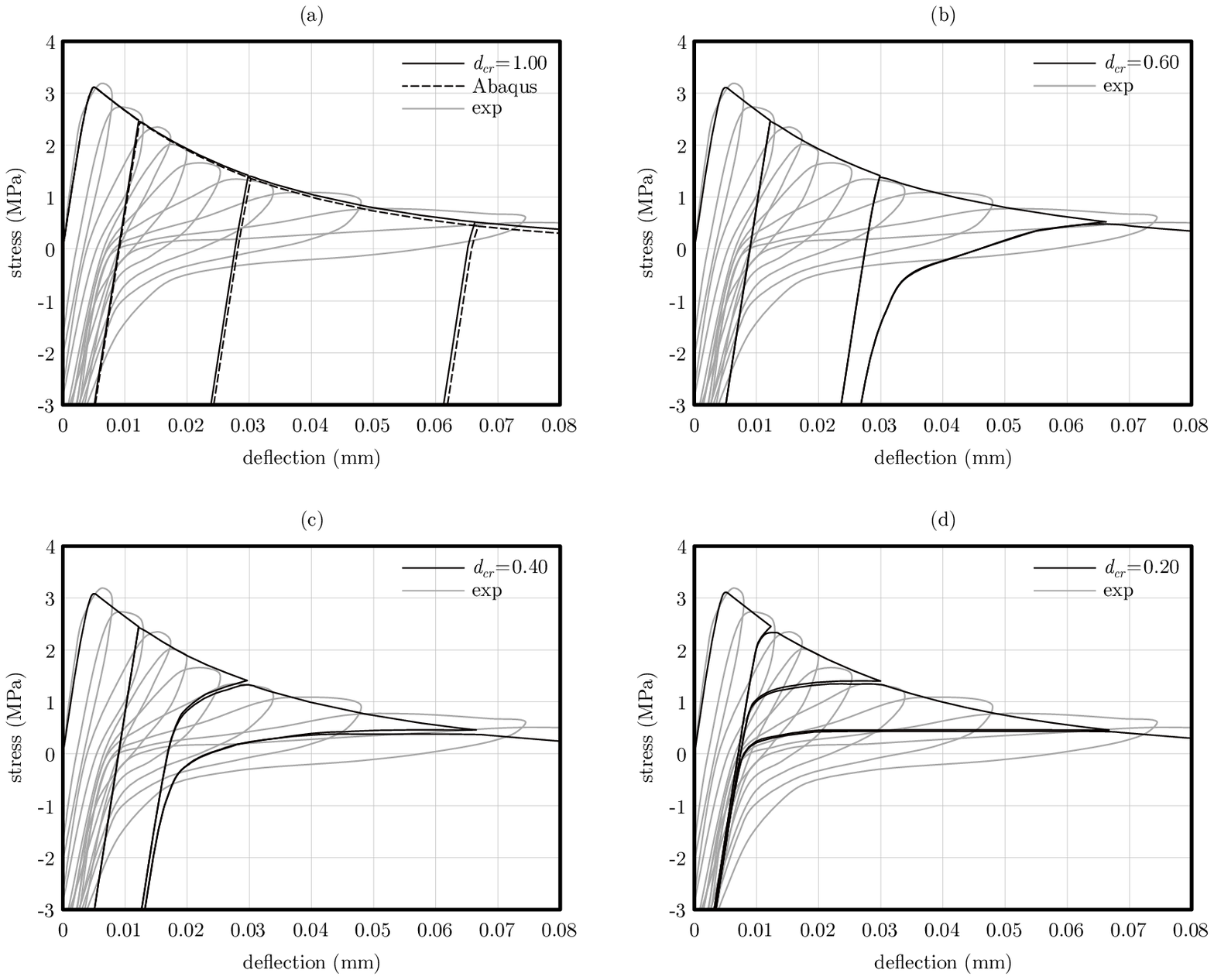}
    \caption{Comparison of the global responses with the experimental data for the full-cycle test: (a) Abaqus model and $d_{cr}=1.00$, (b) $d_{cr}=0.60$, (c) $d_{cr}=0.40$, and (d) $d_{cr}=0.20$.}
    \label{fig:fatigue_curves}
\end{figure*}

The stress versus deflection responses of the numerical models are plotted on the experimental data in Figure \ref{fig:fatigue_curves}. No sign of stiffness degradation can be observed in the unloading branches of the Abaqus model and the $d_{cd}=1.00$ case. On the other hand, by decreasing the value of $d_{cr}$, the contribution of the discontinuity strain field, which reveals itself in the activation of the unilateral effects under less permanent deflection, becomes more pronounced.

\section{Conclusion} \label{sec:conclusion}
This paper introduces a new term to the conventional additive strain decomposition to prevent the excessive unrealistic growth of irrecoverable deformations to reproduce the entire failure process of solids. 

The issue of excessive growth of irrecoverable deformations is rarely addressed (see for example \cite{lee1998plastic, omidi2013continuum}) since it only manifests itself if the unilateral effects resulting from to the closing and reopening of fully formed cracks are of interest. However, this is the case in the area of low cycle fracture/fatigue as caused by earthquakes in civil engineering structures, for example. We demonstrated the problem arising from the mentioned excessive strains which are inherent to other plastic damage models and showed how these strains cause a faulty stress redistribution upon load reversal. 

We called the newly introduced term \textit{discontinuity strain field} since it mimics a discontinuity, yet it is defined at the infinitesimal element level. We demonstrated that introduction of the discontinuity stain field to the conventional additive strain decomposition of the plasticity theory avoids the built up of excessive strains by means of three numerical examples: a pure mode I cracking, a mixed-mode cracking, and a high-intensity low-cycle fatigue test. It is worth mentioning that, although we used the discontinuity strain field to resemble the unilateral effects arising from the collision of the opposing crack faces, it can be used to represent any kind of discontinuity. 

\bigskip

\section{Acknowledgment}
The work was supported by the Alexander von Humboldt Foundation, the Geothermal-Alliance Bavaria (GAB) by the Bavarian State Ministry of Science and the Arts (StMWK) and Deutsche Forschungsgemeinschaft (DFG, Germany) through
the project 414265976 TRR 277 C-01.

\bibliographystyle{ieeetr}
\bibliography{references}

\end{document}